\pdfoutput=1
\PassOptionsToPackage{dvipsnames}{xcolor}
\documentclass[sigconf,nonacm]{acmart}

\usepackage{pvldb}

\usepackage{comment}
\usepackage{endnotes}
\usepackage{graphicx}
\usepackage{tabularx}
\usepackage{arydshln}
\usepackage{multirow, booktabs}
\usepackage{siunitx}
\usepackage{enumerate}
\usepackage{enumitem}
\usepackage{subcaption}
\usepackage{amsmath}
\usepackage[ruled,linesnumbered]{algorithm2e}
\usepackage{listings}
\usepackage{pifont}
\usepackage{makecell}
\usepackage{framed}
\usepackage{hyperref}
\usepackage{xspace}
\usepackage{cleveref}%
\usepackage{tikz}
\usepackage[most]{tcolorbox}
\usepackage{wrapfig}
\usepackage[all]{nowidow}
\usepackage[normalem]{ulem}

\usetikzlibrary{tikzmark, calc, positioning}

\crefname{section}{\S}{\SS}
\crefformat{section}{\S#2#1#3}
\crefformat{subsection}{\S#2#1#3}
\crefformat{subsubsection}{\S#2#1#3}

\setlist{itemsep=0pt,parsep=0pt,topsep=0pt}%
\newcommand*\circled[1]{\tikz[baseline=(char.base)]{
            \node[shape=circle,fill,inner sep=1pt] (char) {\textcolor{white}{#1}};}}

\newcommand{\sys}{\textsf{LooseLeaf}\xspace}

\newcommand{\phead}[1]{\noindent\textbf{#1}}
\newcommand{\ImplBox}[1]{
\vspace{-2pt}
\begin{tcolorbox}[breakable,width=\columnwidth,title={Implications},boxrule=.3mm,colback=white,coltitle=white,left=.5mm, right=.5mm, top=.5mm, bottom=.5mm,boxsep=0.5pt,before skip=3pt,after skip=3pt]
\textit{#1}
\end{tcolorbox}
}

\renewcommand{\shortauthors}{Wu et al.}

\title{LooseLeaf: Enabling Serverless GPUs for Computational Notebooks}

\author{Yusen Wu}
\affiliation{%
  \institution{University of Virginia}
  \city{Charlottesville}
  \state{VA}
  \country{USA}
}
\email{xrn4ub@virginia.edu}

\author{Tingfeng Lan}
\affiliation{%
  \institution{University of Virginia}
  \city{Charlottesville}
  \state{VA}
  \country{USA}
}
\email{erc8gx@virginia.edu}

\author{Benjamin Carver}
\affiliation{%
  \institution{George Mason University}
  \city{Fairfax}
  \state{VA}
  \country{USA}
}
\email{benrcarver@gmail.com}

\author{Haoliang Wang}
\affiliation{%
  \institution{Adobe Research}
  \city{San Jose}
  \state{CA}
  \country{USA}
}
\email{hawang@adobe.com}

\author{Geoffrey Fox}
\affiliation{%
  \institution{University of Virginia}
  \city{Charlottesville}
  \state{VA}
  \country{USA}
}
\email{vxj6mb@virginia.edu}

\author{Yue Cheng}
\affiliation{%
  \institution{University of Virginia}
  \city{Charlottesville}
  \state{VA}
  \country{USA}
}
\email{mrz7dp@virginia.edu}

\begin{document}

\begin{abstract}
Computational notebooks reserve a GPU for the whole session, yet the GPU
computes only while a cell executes and sits idle otherwise.
Serverless execution would release the GPU between cells, but reconstructing
the accumulated notebook state on a fresh worker imposes substantial
cost on every GPU cell.
This paper presents \sys{}, a serverless GPU
platform for interactive notebooks. \sys{} derives the
state and import closures of each cell, the minimal variables and packages
that the cell needs. Our analysis on 5{,}000 real-world notebooks shows that these closures are small, and user-cell editing leaves seconds to prepare them.
\sys{} prepares a GPU worker with these closures while the user
edits the cell. After execution, the worker saves the updated state
within seconds, so the GPU is reclaimable after every cell. \sys{}
starts 95--98\% of GPU cells with no visible delay, matches the
interactivity of a pinned session, and provisions 32--55\% fewer GPU-hours than state-of-the-art notebook platforms.

\end{abstract}

\maketitle
\pagestyle{plain}

\section{Introduction}
\label{sec:intro}

\begin{figure}[t]
  \centering
  \includegraphics[width=0.42\textwidth]{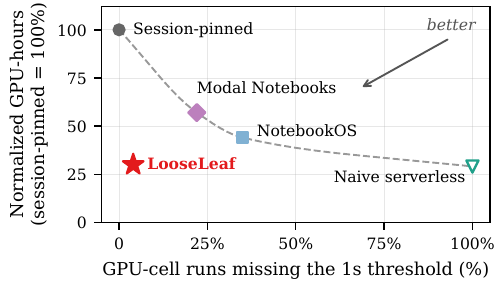}
  \caption{Provisioned GPU-hours against interactivity for \sys{} and
  four baseline approaches on real Kaggle notebooks at 72 concurrent
  sessions (\cref{sec:evaluation}).
  The horizontal axis gives the fraction of GPU-cell runs that miss the
one-second interactive-start threshold.
  The vertical axis gives GPU-hours normalized to session-pinned reservation (e.g., Colab~\cite{colab_web}, Kaggle Notebooks~\cite{kaggle_web}). The baselines trace a cost--interactivity frontier; \sys{} shifts this frontier toward the lower left, approaching the interactivity of session-pinned execution while using fewer GPU-hours than all serverless baselines.}
  \label{fig:cost-interactivity}
\end{figure}

Computational notebooks are the de facto interface for data-driven
programming. A notebook organizes a program into a sequence of
cells. The user runs one cell at a time, and the output
appears as soon as execution finishes. This immediate feedback
suits the exploratory nature of data
analysis and machine learning. Software such as
Jupyter~\cite{jupyter_elpub16} brings this interface to millions of users
across education, research, and
industry~\cite{perkel_nature18, pimentel_msr19}. Every major cloud provider now
offers a commercial notebook
platform~\cite{sagemaker_web, azureml_web, colab_web,kaggle_web}.

Nowadays, AI/ML makes up a growing share of notebook workloads.
A production study reports that interactive deep learning training has
emerged as a dominant notebook use case~\cite{notebookos_asplos26}.
The rapid progress of large language models (LLMs) pushes this trend further.
Open-weight releases put billion-parameter pretrained models in the
hands of individual users~\cite{qwen3_arxiv25, gptoss_arxiv25, gemma4_arxiv26}.
Fine-tuning or validating such a model
for a task takes repeated trials~\cite{dodge_arxiv20, mosbach_iclr21, amershi_icse19}, and notebooks fit this style of
work.
Every one of these steps executes on a GPU. GPU-backed notebooks are
now a standard offering of these platforms~\cite{colab_web,kaggle_web, modalnotebooks_web, databricks_web}.

Platforms such as Colab~\cite{colab_web} and Kaggle~\cite{kaggle_web} bind a GPU to a notebook session for its entire lifetime.
The session acquires the GPU when it starts
and holds the GPU until it ends. A session, however, needs the GPU
for only a small fraction of its lifetime. First, a notebook mixes
GPU cells for training and inference with CPU-only cells for data
preparation and plotting. The GPU does no work while a CPU-only cell
runs. Second, the user pauses between executions to examine the
output, reason about the result, and edit the next cell. This
development time far exceeds the execution time~\cite{notebookos_asplos26}.
The reserved GPU thus computes only while a GPU cell executes and sits idle for the rest of the session. A production study reports that reserved GPUs are idle over 81\% of the time~\cite{notebookos_asplos26}. Session-pinned reservation is therefore wasteful and prohibitively expensive at fleet scale.

A natural remedy is to acquire a GPU only while a GPU cell executes.
Serverless GPUs have recently emerged as a new type of cloud service
and offer exactly this model~\cite{serverlessllm_osdi24,modal_web, runpod_web}. In this
model, code runs as self-contained functions. Each function
invocation is served by a worker process that is created on demand
on a GPU machine and reclaimed once the invocation finishes. A
notebook could thus run each GPU cell as one invocation. The GPU
would be released during the CPU-only cells and the pauses in
between, and provisioning would follow actual demand.

A notebook cell, however, is not a self-contained function.
Cells run in a long-lived kernel process and build on the variables and
packages that earlier cells leave behind. A fresh worker holds none
of this state and needs seconds to minutes per run to re-compute the whole of it, whether by re-running the earlier cells or by loading all previous variables.
Such a start is far from interactive. Prior response-time studies suggest
that responses should arrive within about \emph{one second} for users to
maintain cognitive flow~\cite{miller_afips68,usabilityeng_mk93, Nah01052004}. We therefore use one second as the \emph{interactive-start threshold}: a GPU-cell run misses this threshold if the cell begins executing more than one second after the user presses \emph{Run}.
Serverless GPUs for notebooks thus come down to a question of
session state. A cell must receive the state it needs before it can
run, and the time to provide that state decides whether the cell
starts interactively.

A cell
does \emph{not} need the whole state. A cell that runs
inference, for example, reads the model, the tokenizer, and the
inputs it evaluates. The dataframes and plots that earlier
exploration left behind are irrelevant to it. What a cell needs is
also knowable. Notebook code is organized in cells, so the variables
and packages that a cell references can be read from its code while
the user is still writing it. To verify these properties at scale,
we characterized roughly 5{,}000 real AI/ML notebooks
(\cref{sec:characterization}). The characterization confirms that a
GPU cell typically reads a small fraction of the session state, and
that the packages it imports concentrate in a small common set that
can be loaded once and shared. Notebooks also offer a unique window
to exploit these properties. A user edits a cell for seconds to
minutes before running it. What a GPU cell needs can thus be
prepared during this window, before the user presses Run.

This paper presents \sys{}, a serverless GPU platform built on these
observations. While the user edits a cell, \sys{} derives from the
code the variables and packages that the cell needs. We call these
two sets the state closure and the import closure of the cell.
\sys{} then prepares a serverless GPU worker with these closures ahead
of the run. By the time the user presses Run, the worker already
holds what the cell needs, and the cell starts at once. After the cell executes, the worker saves the state updates (incrementally)
to a shared store within seconds, so the GPU is
reclaimable right after every cell. \sys{} therefore provisions GPUs
by actual demand.
\Cref{fig:cost-interactivity} summarizes the key results on real Kaggle
notebooks. Across all notebook solutions, \sys{} keeps user-visible start
latency below one second at p95: with 72 sessions, more than 95\% of GPU cells begin execution within one second after the user presses \emph{Run}.

In summary, this paper makes the following contributions:
\begin{itemize}
  \item We characterize roughly 5{,}000 real-world AI/ML
  notebooks and make three observations that motivate
  \sys{}. %
  \item We introduce per-cell state and import closures, two
  abstractions that reduce the preparation of a GPU cell from the
  whole session to what the cell needs. %
  \item
  We design a fast checkpoint store for the session state
  of GPU notebooks. On top of it, we build \sys{}, a serverless GPU platform that
  prepares a GPU worker for each GPU cell during editing.
  \item
  We evaluate \sys{} on real Kaggle notebooks. Across loads, \sys{}
  provisions 43--70\% fewer GPU-hours than session-pinned
  reservation and 32--55\% fewer than the state-of-the-art
  NotebookOS.

\end{itemize}

\section{Background and Motivation}

\subsection{Interactive Notebooks and Serverless GPUs}
\label{sec:notebooks}

Notebooks
are widely used in data-driven programming, particularly in AI/ML workflows. They provide a user-friendly programming interface where a program is divided into cells, each
containing a block of code. The intermediate outputs can be inspected immediately after cell execution.
This immediate feedback enables an iterative workflow in which users write and execute code in a cell, examine the results, and then proceed to the next. Users can also revisit
and re-execute earlier cells as they refine their analysis.

To support this interactive workflow, each notebook runs within a
session backed by a long-running process called the kernel. All cells execute in this kernel and share an implicit namespace: the variables and packages one cell creates persist and become available to later cells without re-execution. Each cell thus inherits the state left by earlier cells and adds its own, so the namespace grows as the session proceeds.

With the development of AI, there is an increasing demand to run notebooks on GPU-equipped machines for deep learning applications (e.g., model fine-tuning, hyperparameter tuning, and inference)~\cite{notebookos_asplos26}.
In existing notebook services such as Colab~\cite{colab_web} and Kaggle~\cite{kaggle_web}, a notebook session retains the GPU throughout its entire lifetime. However, a typical notebook contains both CPU-only cells and cells that require GPU resources, and users often pause between cell executions to inspect outputs, reason about results, or step away. As a result, the GPU is needed by only a subset of cells for execution and sits idle for the rest of the time. %

In recent years, serverless GPUs have emerged as a cloud service
model~\cite{serverlessllm_osdi24, modal_web,runpod_web}. Users submit code as
\emph{functions} rather than rent GPU machines. Each function execution,
or \emph{invocation}, runs on a \emph{serverless GPU worker}: a process
created on demand on a GPU host and reclaimed after the invocation and any
keep-alive interval~\cite{modal_keepalive,serverlesswild_atc20}. Users are typically billed for invocation execution rather than full machine
reservation. A notebook can therefore run each GPU cell as one invocation,
holding a GPU only during execution and releasing it during CPU-only cells
and inter-cell pauses. Serverless GPUs are thus a natural fit for notebook
workloads.

\subsection{State and Import Closures}
\label{sec:closures}

\begin{figure}[t]
  \centering
  \includegraphics[width=\columnwidth]{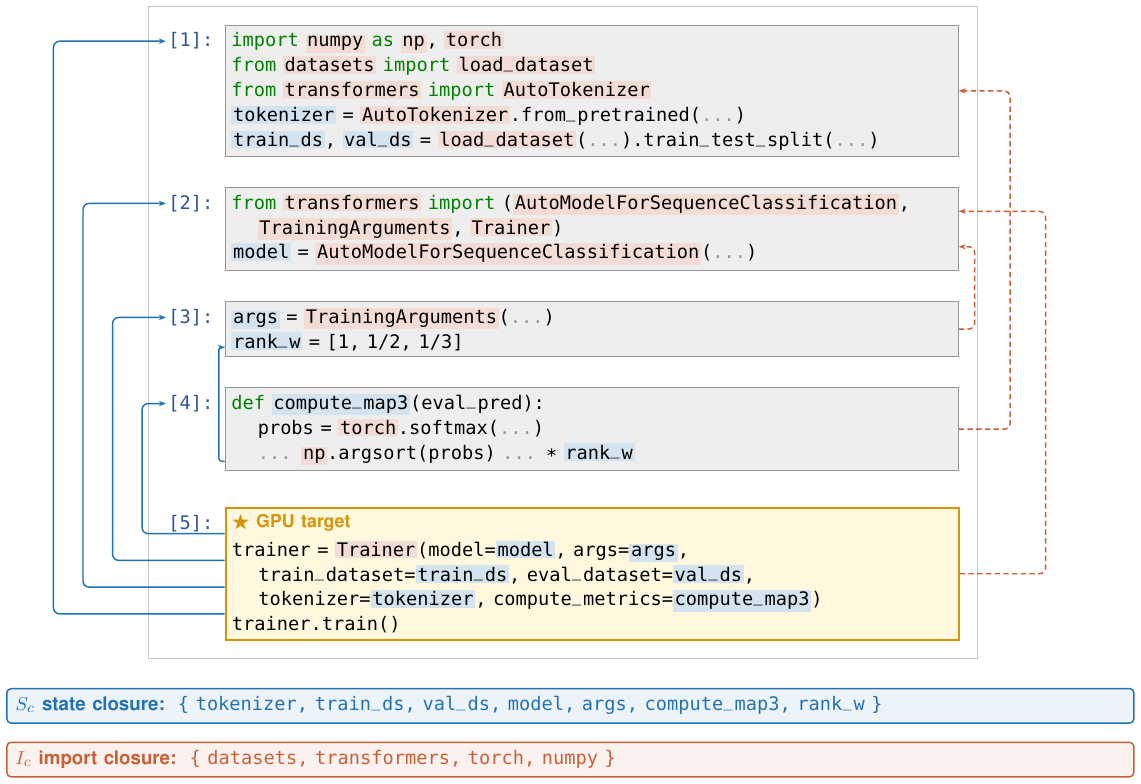}
  \caption{Dependency graph of a five-cell model fine-tuning notebook.
  Cell~5 is the GPU target. Solid arrows are state dependencies, and
  dashed arrows are import dependencies. Both traces are transitive.
  The boxes at the bottom show the resulting closures $S_c$ and $I_c$.}
  \label{fig:dag-closure}
\end{figure}

\begin{figure*}[!t]
  \centering
  \includegraphics[width=0.90\textwidth]{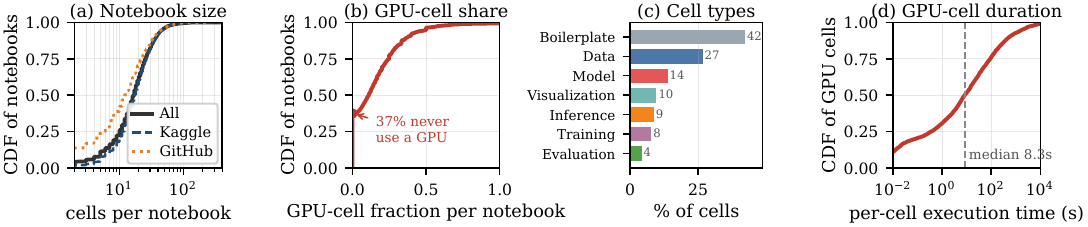}
  \caption{General statistics of the collected notebook corpus.
  (a)~CDF of cells per notebook; a point $(x, y)$ means a fraction $y$
  of notebooks contain at most $x$ cells. (b)~CDF of the per-notebook
  fraction of GPU cells; the jump at zero corresponds to notebooks that
  never use a GPU. (c)~Cell type distribution. (d)~CDF of
  per-cell execution time over GPU cells, excluding definition-only
  cells; the dashed line marks the median.}
  \label{fig:general}
\end{figure*}

\phead{The stateful--stateless tension.}
Although serverless GPUs are a natural fit for notebook workloads,
applying them introduces a fundamental tension.

Notebook cells communicate through an implicit shared namespace,
where each cell reads objects defined by prior cells and leaves its own
updated objects for subsequent ones. The notebook session is therefore
stateful. By contrast, a serverless invocation is stateless~\cite{berkeleyview_arxiv19, onestepback_cidr19}.
To execute
a GPU cell as a serverless function on a newly created runtime, the
runtime must
1)~prepare the execution environment by initializing the
CUDA runtime and importing all packages that the kernel had previously
loaded,
and 2)~restore all variables accumulated by prior cells.
We call the
end-to-end cost of making the worker ready to execute the cell its
\emph{preparation latency}. This cost includes both serverless cold-start
overhead and notebook-state restoration, and is prohibitively high when all
preparation begins after the user presses ``Run.''

\phead{Dependency-guided closures.}
The key observation is that notebooks expose program structure at cell granularity. By statically analyzing each cell, the runtime identifies the variables it defines and references, and these definition--use (def-use) relationships form a dependency graph over cells. An edge connects a reference to its definition. Because a name may be defined by more than one cell and cells may run in any order, the edge points to the definer that executed most recently. \sys{} uses this graph to compute each cell's execution requirements precisely. For a target cell, we collect its referenced variables and perform a BFS over the graph to obtain the minimal state required to run it, and apply the same analysis to imported packages to obtain its minimal import requirement. This lets us model each cell as a serverless function
    $F_c(S_c, I_c) \rightarrow (S'_c, O_c)$,
where the \emph{state closure} $S_c$ is the minimal set of live variables the cell must read, the \emph{import closure} $I_c$ is the minimal set of packages it must import, $S'_c$ is the updated state, and $O_c$ is the output.
Static analysis can determine both closures exactly, because every
variable and package that a cell uses is named in its code or in the
body of a function that it references. The only exception is code that reads the
namespace dynamically through constructs such as \texttt{exec} or
\texttt{globals()}, which notebook code rarely uses.
The state closure and import closure are the two abstractions on which the rest of the paper builds.

\definecolor{statebg}{HTML}{EAF4FF}
\definecolor{statefg}{HTML}{2F75B5}
\definecolor{importbg}{HTML}{FDEDEA}
\definecolor{importfg}{HTML}{D9653B}

\newcommand{\statehl}[1]{%
  \begingroup
  \setlength{\fboxsep}{0pt}%
  \colorbox{statebg}{%
    \hspace{1.5pt}%
    \textcolor{statefg}{\strut #1}%
    \hspace{1.5pt}}%
  \endgroup
}

\newcommand{\importhl}[1]{%
  \begingroup
  \setlength{\fboxsep}{0pt}%
  \colorbox{importbg}{%
    \hspace{1.5pt}%
    \textcolor{importfg}{\strut #1}%
    \hspace{1.5pt}}%
  \endgroup
}

\Cref{fig:dag-closure} illustrates this analysis on a five-cell
model fine-tuning notebook. Cell~5 is the GPU target cell that
constructs a \texttt{Trainer} object and launches model training. By
tracing the state dependencies (solid arrows), the runtime determines
that Cell~5 directly references six variables defined by the preceding four
cells: \statehl{\texttt{model}}, \statehl{\texttt{args}}, \statehl{\texttt{train\_ds}},
\statehl{\texttt{val\_ds}}, \statehl{\texttt{tokenizer}}, and \statehl{\texttt{compute\_map3}}. The
trace also follows references inside function bodies.
\statehl{\texttt{compute\_map3}} reads the list \statehl{\texttt{rank\_w}}  defined in
Cell~3, so \statehl{\texttt{rank\_w}} joins the closure even though Cell~5
never names it. These
seven variables form the state closure $S_c$. In parallel, the runtime
traces import dependencies (dashed arrows) to identify the packages that
the cell and its transitive dependencies require. For example, the
function \statehl{\texttt{compute\_map3}} defined in Cell~4 calls
\texttt{torch.softmax} and \texttt{np.argsort},
so \importhl{\texttt{torch}} and
\importhl{\texttt{numpy}} must be available even though Cell~5 does not import
them directly. Together with \importhl{\texttt{datasets}} and \importhl{\texttt{transformers}},
these four packages form the import closure $I_c$. A
GPU worker
that receives only these two closures can execute Cell~5 without
restoring the full session namespace.

\subsection{Dataset Collection and Statistics}
\label{sec:dataset}

The closures defined above narrow the two sources of preparation
latency (\S\ref{sec:closures}) to per-cell scope: the import closure
reduces environment preparation from all session packages to the
minimal set a cell requires, and the state closure reduces variable
restoration from the full namespace to the minimal set a cell
references.
Although the closures reduce the scope, this preparation
work remains on the critical path if it begins only after the user
requests execution. To understand the practical magnitude of each
component, we characterize real-world AI/ML notebook workloads.

\begin{wraptable}{r}{0.55\columnwidth}
  \caption{Task composition of the notebook corpus.}
  \label{tab:tasks}
  \scalebox{0.75}{
  \begin{tabular}{l r r}
    \toprule
    \textbf{Task type} & \textbf{Notebooks} & \textbf{Share} \\
    \midrule
    Training                & 2{,}597 & 52\% \\
    Fine-tuning (PEFT)      & 1{,}016 & 20\% \\
    Inference               &   988 & 20\% \\
    Data analysis and other &   403 &  8\% \\
    \midrule
    Total                   & 5{,}004 & 100\% \\
    \bottomrule
  \end{tabular}
  } %
\end{wraptable}
We collected
5,004 notebooks from
Kaggle~\cite{kaggle_web} and public GitHub
repositories~\cite{github_web} by searching
for AI/ML and GPU keywords such as \texttt{llama}, \texttt{lora}, and
\texttt{pytorch}.
To improve labeling robustness, we independently applied two large language models (LLMs), GPT-oss-120B~\cite{gptoss_arxiv25} and Grok-code-fast-1~\cite{grokcodefast_web}, to classify the device requirement of each cell.
A cell was labeled as
GPU-requiring when either model assigned a GPU to it, and was
labeled as CPU-only otherwise.
On the 5{,}004 notebooks processed by both models, the models agreed on
94.5\% of the cell-level labels. Cohen's $\kappa$ was 0.75
~\cite{kappa_epm60}, corresponding to substantial agreement under the
Landis--Koch interpretation~\cite{landiskoch_biometrics77}.

\begin{figure*}[!t]
  \centering
  \includegraphics[width=0.90\textwidth]{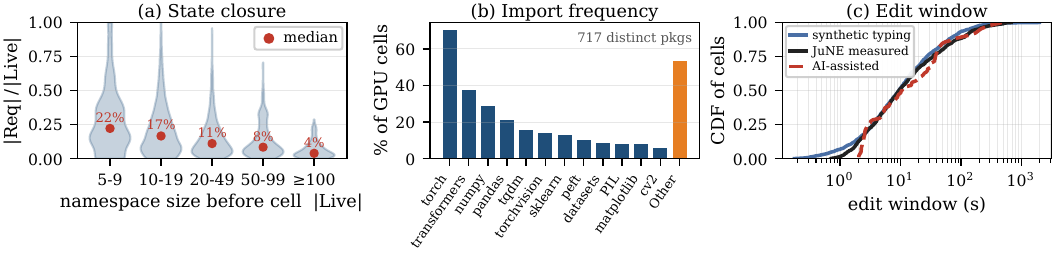}
  \caption{Three properties of the corpus. (a)~State-closure ratio
  against the number of live variables before each GPU cell. (b)~Package frequency over GPU cells. (c)~Per-cell
  editing-window distributions for manual typing and AI-assisted
  editing.}
  \label{fig:obs}
\end{figure*}

Table~\ref{tab:tasks} summarizes the composition of the corpus. Most
notebooks perform model training, fine-tuning, or inference, and
8\% are devoted to data analysis or other tasks.
Figure~\ref{fig:general}(a) shows the distribution of the number
of cells per notebook. A typical notebook contains 17 cells. Notebooks
are thus fairly long, which suggests that a session can accumulate
substantial state as it advances through the cells.
Figure~\ref{fig:general}(b) shows the fraction of GPU cells within
each notebook. A GPU is never used in 37\% of the notebooks, and among
the notebooks that use one, GPU cells account for a median of 20\% of
all cells. Figure~\ref{fig:general}(c) gives a consistent view by cell
type. Boilerplate, data handling, and visualization account for most
cells, while model, training, and inference cells form a minority.
Both views confirm the mismatch described in \cref{sec:notebooks}. A
GPU bound to a whole session serves only a small fraction of its
cells. Figure~\ref{fig:general}(d) shows the execution time of GPU
cells. The distribution spans from milliseconds to hours, and the
median GPU cell executes for only about eight seconds. Most GPU cells
are thus short steps that demand the same interactivity as the rest of
the session. The preparation latency of a GPU invocation must
therefore be small.

\subsection{Dataset Characterization}
\label{sec:characterization}

We then construct the dependency graph and compute the state closure and import closure for each cell.
From this characterization, we derive the following three important findings:
(1)~the state a GPU cell needs is small (\cref{sec:obs1});
(2)~the packages it needs are concentrated (\cref{sec:obs2});
and (3)~editing exposes a window in which the preparation work
can be hidden (\cref{sec:obs3}).

\subsubsection{Observation 1: A GPU Cell's State Closure Is Small}
\label{sec:obs1}

For each GPU cell $c$, we compare the state required by the cell with the named state accumulated before the cell executes. We define $\mathrm{Live}(c)$ as the set of data variables available before $c$ executes, and $\mathrm{Req}(c)$ as the cell's state closure $S_c$, i.e., the variables it must read. The ratio $|\mathrm{Req}(c)| / |\mathrm{Live}(c)|$ measures the fraction of the live namespace that must be materialized for the invocation.

Figure~\ref{fig:obs}(a) plots this ratio against the
namespace size before each GPU cell. The median ratio decreases as the
live namespace grows. Among GPU cells with 5 to 9 live variables, the
median cell reads 22\% of its live variables. For cells with 20 to 49
live variables, the median ratio drops to 11\%, and for cells with 100
or more live variables, it falls to 4\%. These results show that the
cost of restoring a full namespace grows with the accumulated session
state, while the state required by a GPU cell grows much more slowly.
This trend suggests that instead of reconstructing a full kernel
namespace for every GPU invocation, the runtime should restore only the
dependency-guided state closure of the target cell.

\ImplBox{The runtime should restore notebook state at variable granularity using the dependency-guided closure of the target cell.}

\subsubsection{Observation 2: GPU Cells Have a Head-and-Tail Package Distribution}
\label{sec:obs2}

For each GPU cell, we compute its import closure, which is the set of packages that must be available before the cell runs on a GPU.
We aggregate these packages across GPU cells to obtain the package-frequency distribution in Figure~\ref{fig:obs}(b).

The distribution has a small common head and a long tail. Common AI/ML and data packages dominate the head. \texttt{torch} appears in 71\% of GPU cells, followed by \texttt{transformers}, \texttt{numpy}, \texttt{pandas}, \texttt{tqdm}, \texttt{torchvision}, and \texttt{sklearn}. At the same time, the corpus contains 717 distinct packages, and 53.5\% of GPU cells use at least one package outside the top 12.
This distribution exposes a tradeoff in the preparation of GPU
workers. Starting each worker from an empty process pays the same
import cost for the head packages on every invocation. Preloading a
fixed set of packages avoids this repeated cost, but the long tail
means that any fixed set either misses the packages of many cells or
loads hundreds of packages that most cells never use.

This head-and-tail structure is similar to the package distribution reported by SOCK for Python serverless workloads~\cite{sock_atc2018}.
SOCK addresses the same tradeoff with zygote processes~\cite{zygote_web}: workers fork from a process with the common packages already initialized and import the remaining tail packages after the fork. However, SOCK targets CPU-based serverless functions, while our workload requires GPU execution.  The same mechanism cannot be applied directly because a forked child
process inherits the CUDA context handles of the parent (see \cref{sec:design-env}), but the underlying GPU driver resources are not duplicated by
\texttt{fork}~\cite{cudafork_web}.
Still, the similar package distribution suggests that zygote-style process reuse is a promising basis for reducing package setup cost, as long as it handles the CUDA runtime correctly.

\ImplBox{The runtime should reduce environment setup time with a GPU-aware import cache.}

\subsubsection{Observation 3: Editing Provides a Short Preparation Window}
\label{sec:obs3}

The first two observations reduce the amount of state and environment work required by a GPU invocation. However, this work still increases interactive latency if it starts only after the user requests execution. Notebook editing provides an opportunity to move this preparation earlier. While a user edits a cell, the runtime can keep the dependency graph and the cell's closures up to date. It can then start speculative preparation before execution begins.

To estimate the size of this window, we used a typing-speed model from a large-scale keystroke study~\cite{typing_chi18}. We validated this estimate with JuNE~\cite{june_jss26}, a Jupyter notebook development trace dataset that records timestamped cell edits and executions.
The JuNE dataset contains logs of code evolution in Jupyter notebooks, comprising over 100 hours of execution logs from 20 participants solving data science tasks.
Figure~\ref{fig:obs}(c) compares the synthetic typing-time distribution with measured pure-typing intervals. The two distributions closely match, with median times of 9.4s and 9.5s, respectively. This result shows that editing provides a real preparation window, but the typical window is only on the order of seconds.

As LLMs mature, users increasingly rely
on them to assist programming. In a notebook, the user asks an AI
assistant for code and pastes the suggestion into a cell. Pasting takes
far less time than typing, so this interaction could shrink the editing
window.
We examined this concern with the interaction log released by
DSPM~\cite{dspm_chi26}, a study of how students use an LLM assistant~\cite{gemini_arxiv23} for data
analysis in computational notebooks.
From the timestamped edit and execution events in the log,
we measured the interval between each
paste of generated code and the next execution. The distribution of
this interval closely tracks the typing window, with a median of 10.7s
(\Cref{fig:obs}(c)). The user thus still spends seconds reading
and adapting the generated code before running it, and the editing
window remains.

\ImplBox{The runtime should prepare the GPU worker while the user edits the cell, and complete the preparation within the few-second editing window before the user requests execution, thereby keeping preparation off the interactive path.}

\section{Design}
\label{sec:design}

\begin{figure}[t]
  \centering
  \includegraphics[width=0.38\textwidth]{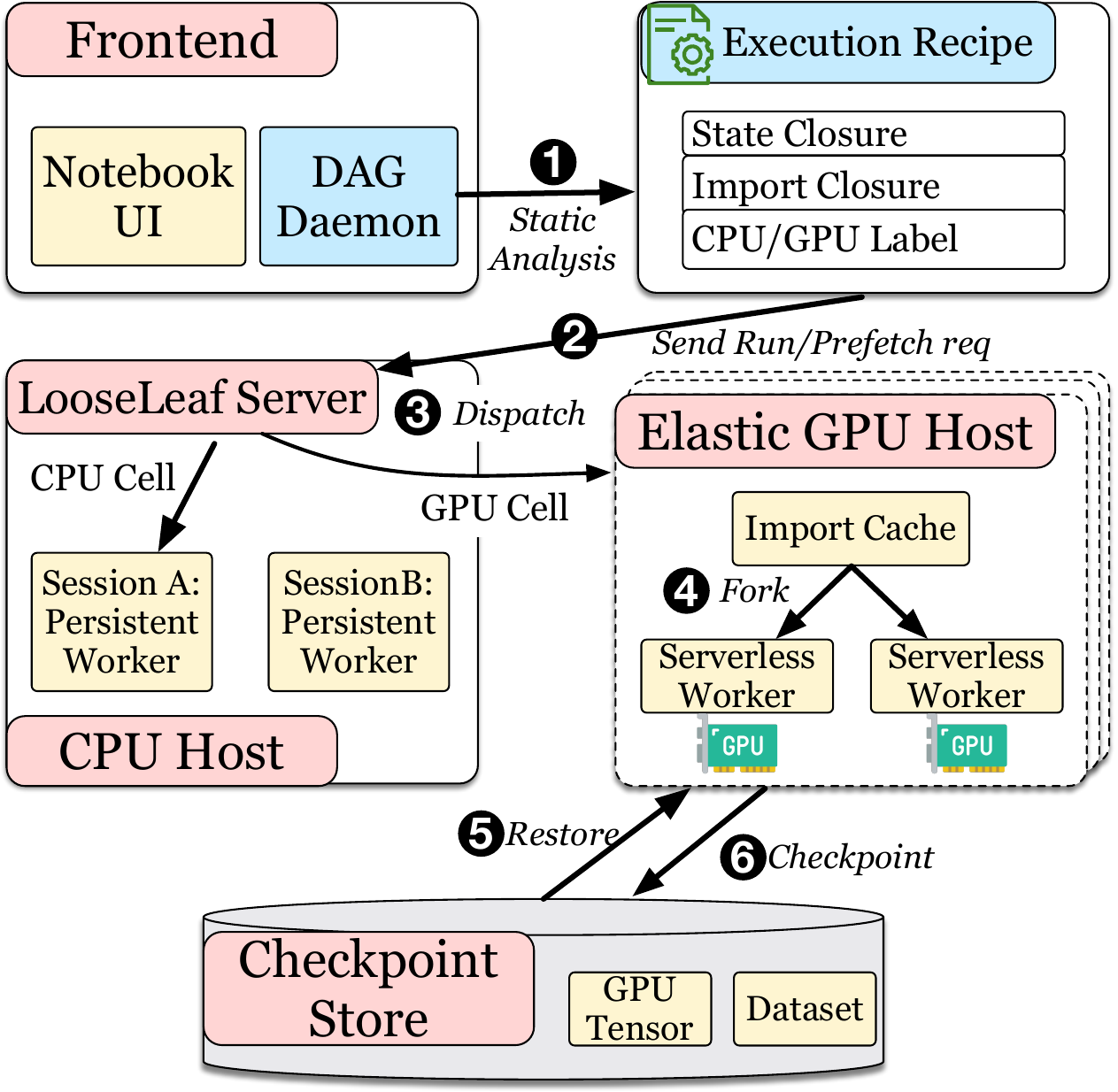}
  \caption{Architecture of \sys{} and the path of a single request through it.
  As the user edits, the DAG daemon distills each cell into an execution recipe
  that holds the state closure, the import closure, and a CPU or GPU label. A CPU cell runs on the
  persistent worker of its session. A GPU cell runs on a worker forked from
  the import cache on a GPU host. Workers exchange notebook state through the
  checkpoint store. Circled numbers mark the steps described in the text.}
  \label{fig:overview}
\end{figure}

\subsection{\sys{} Overview}
\label{sec:overview}

Based on the observations from \cref{sec:characterization}, we design a serverless GPU platform for computational notebooks called \sys{}.
For a GPU cell, \sys{} runs it as a serverless invocation and
provisions a GPU worker on demand using
the state and import closures of the cell.
\sys{} keeps the session state in a checkpoint store shared across hosts (\cref{sec:design-state}). A GPU worker restores the variables of the cell from the store and saves the updated variables back after execution. Therefore, \sys{} can reclaim GPU between cells. The remainder of this subsection describes the components of \sys{} and the path of a GPU cell through them.

\phead{Architecture.}
As shown in Figure~\ref{fig:overview}, \sys{} comprises five
components. The \emph{frontend} hosts the notebook UI and a DAG daemon that
keeps the dependency graph current as the user edits and derives from it an
execution recipe for each cell, which holds the state and import closures of the cell
together with a CPU or GPU label. The \emph{server} is the control plane that dispatches cells and manages
workers. It does not require colocated GPUs and may run on the user's local machine or on a remote host, provided that it can communicate with the frontend, worker hosts, and checkpoint store.
A cell runs on one of two \emph{workers},
a persistent CPU worker that hosts the notebook kernel of a session or an
ephemeral GPU worker that runs a single GPU cell. The \emph{import cache} is a
per-host process pool from which each GPU worker is forked to reduce package
import time.
The \emph{checkpoint store} holds notebook variables and moves them across workers and hosts.

\phead{Workflow.}
A GPU cell passes through \sys{} in two phases. While the user edits the cell,
the DAG daemon updates its recipe by static analysis (\circled{1}). Once the
cell is labeled as a GPU cell, the daemon sends the recipe to the server as a
prefetch request (\circled{2}). The server dispatches the request (\circled{3})
and acquires a GPU worker for the session, which is either a warm worker or a
new worker forked from the import cache (\circled{4}).
The fork provides the packages that the cache has already loaded. The worker then imports the remaining packages of the import closure and restores the variables of the state closure from the checkpoint store (\circled{5}).

When the user runs the cell, the run request carries the
closures computed from the submitted code of the cell,
and the worker completes any remaining preparation before it
executes.
After execution, the worker writes the updated variables back to the
store (\circled{6}) so that later cells can restore them.
The worker then stays warm for the session until \sys{} reclaims it under the eviction policy (\cref{sec:design-spec}).
The following subsections detail these mechanisms.

\subsection{Dependency-Guided Checkpoint / Restore}
\label{sec:design-state}

A GPU worker must hold the variables in the state closure of the cell before the cell can run. \cref{sec:obs1} shows that a GPU cell reads only a small fraction of the
live namespace.
\sys{} exploits this property by restoring
state at the granularity of individual variables. For a target cell, \sys{}
computes the state closure based on the dependency graph and restores only those variables. Between cells, these variables live in the checkpoint store.
The rest of this subsection describes how a variable moves into the store and back.

\phead{The problem with per-variable pickle.} To restore variable-level state,
the straightforward approach is to
serialize each variable with ordinary pickle~\cite{pickle_web}.
Pickle encodes the
variable into a self-contained stream carrying two kinds of content
for every object, the structure and the payload. The structure describes how to rebuild the object. The
payload is the raw data bytes that the object carries.
With one stream per variable, the runtime can save and restore each variable independently, which is the granularity that the state closure requires. However,
this approach is both incorrect and slow.

The approach is \emph{incorrect} because variables share objects. For example, an
optimizer maintains parameter groups that refer to model parameters.
However, pickle detects sharing only within the single stream that it writes.
This is because pickle detects sharing through a memo that records
every object written to the stream, and the memo is discarded when the
stream ends. The model and the optimizer are serialized into two separate streams, so each stream carries a private copy of the shared parameters. Restore then
materializes the two copies as distinct objects. As a result, a training step updates
the copy held by the optimizer while the model computes with the other
copy.
The same failure arises for tensor views~\cite{tensorview_web}.
Slicing and reshaping can produce a view, which is a distinct tensor object that shares the storage of its base tensor. If the view and the base tensor are serialized into two
separate streams, each stream again carries a private copy of that storage.

The approach is also \emph{slow}, because a pickle stream must
materialize every payload byte inside it. Saving a variable copies
the payload of each object at least twice, and restoring copies it
back out. A GPU tensor pays two more transfers because its bytes
cross the PCIe bus at each end. All of these copies run sequentially
in one thread. The serialization cost of a variable therefore grows
with its payload size. Unfortunately, almost all state bytes in a
notebook sit in such payloads. On the notebooks used in our
evaluation, GPU tensors hold 95.0\% of all saved state bytes, and CPU
buffers such as dataset and array payloads hold most of the
remainder. The other objects are small control values and together
hold under one percent. The cost of pickling a closure is thus the
cost of
moving its payload bytes. The sharing problem makes this even worse.
Pickle cannot recognize a shared tensor across separate streams, so
every stream that reaches the tensor moves its bytes again.

\begin{figure}[t]
  \centering
  \includegraphics[width=0.42\textwidth]{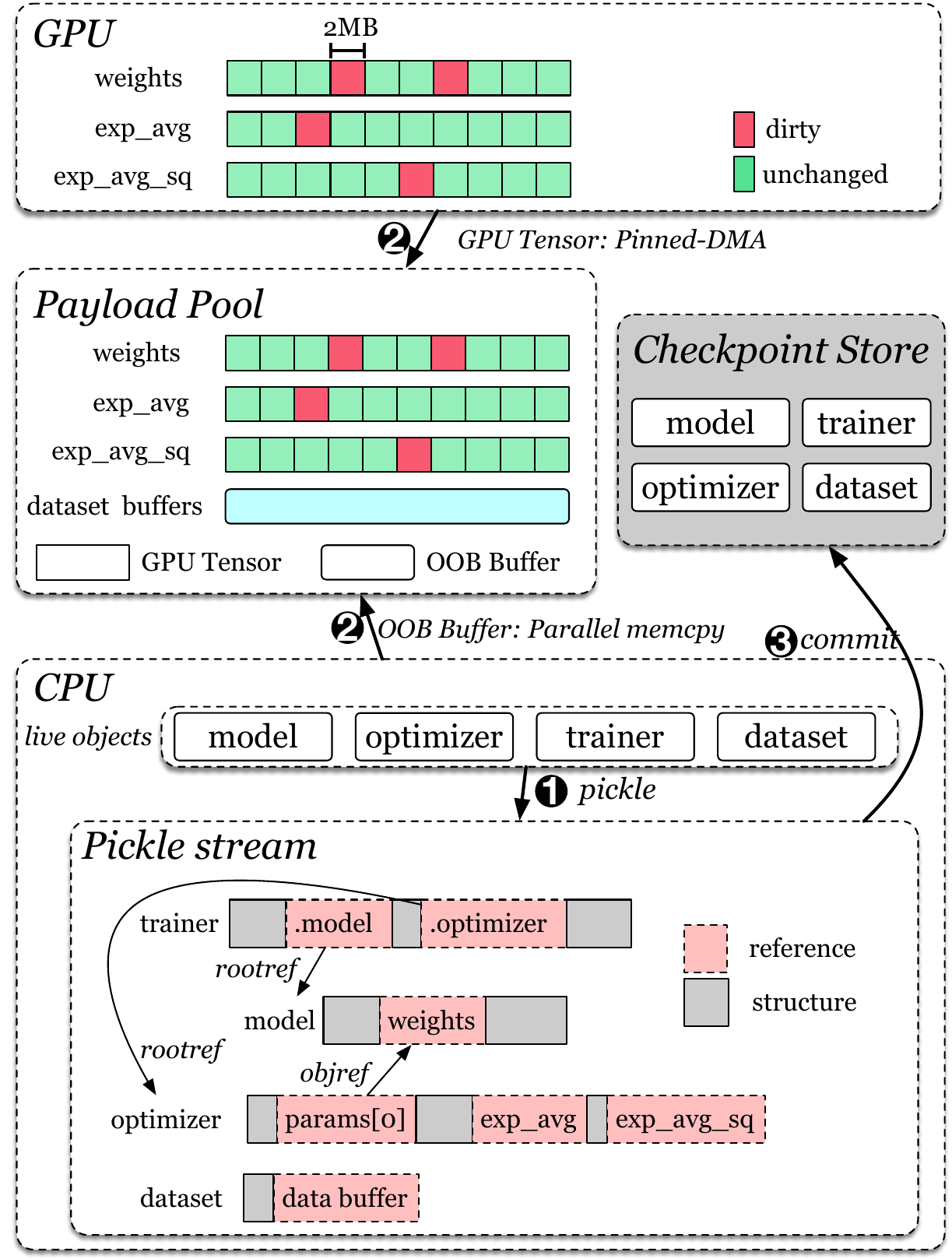}
  \caption{Checkpointing a cell's state. Pickling captures the object
  structure and makes cross-variable sharing explicit through
  \texttt{RootRef} and \texttt{ObjRef} references (\protect\circled{1}).
  Large payloads leave the stream for a shared pool
  (\protect\circled{2}), and the structure blobs commit to the
  checkpoint store (\protect\circled{3}).}
  \label{fig:checkpoint}
\end{figure}

Therefore, variable-level restore requires both a mechanism to preserve
sharing across variables and a mechanism to keep the payload bytes out of
the pickle stream. Once the payloads leave the stream, they can move on
fast bulk paths. The stream shrinks to the object structure and the
small control values that pickle serializes well.
\Cref{fig:checkpoint} illustrates the two mechanisms on a training closure.

\phead{Preserving cross-variable sharing.}
\sys{} preserves sharing by encoding cross-variable references explicitly.
To detect sharing across variables, \sys{} retains the memo of
every variable after serialization. The retained memo keeps
holding the objects that it records, so the identity of a recorded
object cannot be reused by another object.
While serializing a variable, the serializer
checks each object that it reaches
against the retained memos of the other variables of the session. A
match shows that another variable has already serialized the object.
In this case, \sys{} emits a reference instead of serializing another
copy of the object. A \texttt{RootRef} names another top-level
variable, whereas an \texttt{ObjRef} names an object inside another serialized variable, such as
a model parameter or a buffer. In \Cref{fig:checkpoint}, \texttt{trainer}
holds a \texttt{RootRef} to each of the top-level variables
\texttt{model} and \texttt{optimizer}, and the parameter entry inside
\texttt{optimizer} is an \texttt{ObjRef} to the \texttt{weights}
serialized under \texttt{model}. Immutable objects are
serialized by value because sharing is unobservable for them. This
mechanism allows variables to be serialized separately while preserving
the sharing between them.

\phead{Externalizing large payloads.}
To keep the payload bytes out of the stream, \sys{} intercepts the
serialization of GPU tensors and CPU buffers. These two classes hold
nearly all of the saved state bytes. When the pickler reaches such a payload,
\sys{} moves the payload bytes out of the stream into a shared pool on the
host. The pool is backed by huge pages~\cite{hugetlb_web}, which reduces page fault time.
In the stream, a reference to this pool location takes the place of the
payload. The payload of a storage is written once, by the variable
that first saves it. Any other variable that reaches the same storage
references this pool location and writes nothing. The stream therefore keeps only the object structure and the
references. Each payload class enters the pool on the bulk path that fits
it. For a GPU tensor, the worker registers the pool regions with CUDA, so
the tensor storage moves between device memory and the pool in one pinned
DMA transfer.
For a dataset or array buffer, \sys{} uses pickle protocol~5's
out-of-band (OOB) buffer mechanism~\cite{pep574_web} to separate the
buffer payload from the pickle stream. \sys{} copies the resulting
buffer into the payload pool using parallel \texttt{memcpy}.
With every payload outside the stream, the pickle stream of each variable
shrinks to a small blob of structure and references
(\Cref{fig:checkpoint}), and the memo that \sys{} retains for
sharing detection covers only this structure. Each variable then commits this blob to the checkpoint store. The
store keeps one entry per variable of the session and records on which
host the payload bytes of the session reside. When the session moves to
another host, \sys{} copies the payload bytes from the old pool into the
pool of the new host (\cref{sec:design-spec}).

\phead{Restore.}
At restore time, \sys{} starts from the state closure of the target cell. It unpickles the variables of the closure in dependency order. When
unpickling reaches a \texttt{RootRef} or an \texttt{ObjRef}, \sys{} restores
the referenced object and reconnects the reference to it, so a shared object
is restored once. In \Cref{fig:checkpoint}, restoring \texttt{trainer} pulls
in \texttt{model} and \texttt{optimizer} through its \texttt{RootRef}s, and
the \texttt{ObjRef} inside \texttt{optimizer} reconnects to the
\texttt{weights} restored under \texttt{model}. For a payload reference, \sys{}
fetches the payload bytes from the shared pool. The runtime therefore restores the DAG-selected variables together with
their shared objects and reachable payload bytes. This restore is minimal, and the same bound applies to GPU memory. Only the tensors inside this
reachable set move onto the GPU. When the session holds both a model and an
optimizer and the target cell reads only the model, the closure names the
model alone. \sys{} moves the \texttt{weights} onto the GPU and leaves the
optimizer state in the checkpoint store.

\phead{Incremental save.}
After a cell executes, the worker saves the updated state
$S'_c$. Besides the variables that the cell defines, $S'_c$ includes
variables in the state closure that the cell changes in place, such as
by appending to a list. Static analysis cannot see such changes, so
the worker re-serializes every variable in the state closure and every
variable reachable from it through references. The re-serialization
rebuilds only the structure blobs, which are small because the
payloads live in the pool.
The payload bytes are updated by dirty tracking. For
GPU tensors, \sys{} tracks dirty regions at 2\,MB granularity. After a cell
updates tensor values, \sys{} refreshes only the dirty regions in the pool
instead of rewriting the full tensor payload (the dirty regions of \texttt{weights} and the optimizer state in \Cref{fig:checkpoint}).
Because a variable is serialized again at every save that
reaches it, the positions of its objects in the stream can change. An
\texttt{ObjRef} therefore carries a permanent id of the object. The
store maps the id to the current location of the object, and the owner
updates this location at each save.

\phead{State coherence across workers.}
The persistent CPU worker and the ephemeral GPU workers exchange
variables only through the checkpoint store. After a cell executes,
the worker saves the variables that the cell updated, and the next
cell reads them from the store on any worker. The store keeps a
version for every variable and advances it whenever a save changes the
variable. A worker that already holds a copy of the variable refreshes
the copy when the version has advanced.

\subsection{GPU-Aware Import Cache}
\label{sec:design-env}

Restoring the state closure materializes objects whose classes are
defined in packages, and executing the cell calls packages directly.
Before the restore begins, the worker therefore needs to initialize
CUDA and import the packages of the import closure.
As \cref{sec:obs2} shows, GPU cells follow a head-and-tail package
distribution. Many cells use the same common AI/ML and data packages,
while a long
tail of packages appears in only a small fraction of cells. If each worker
imports packages from scratch, the common head causes worker creation to pay the
same import cost repeatedly.
\sys{} reuses redundant import packages for
GPU cells through an import cache.
A worker forked from the import
cache inherits the loaded modules through copy-on-write and imports only the
missing packages after the fork.

\phead{The CUDA fork-safety problem.}
However, this fork-based method cannot be applied to GPU cells, because CUDA
runtime state is not fork-safe~\cite{cudafork_web}.
A CUDA context is bound to the process
that initializes it and cannot be used by a forked child, so a process that has initialized CUDA cannot serve as a parent for GPU workers. Therefore, every fork parent in the import cache must remain
\emph{CUDA-clean} and exclude packages whose imports touch CUDA.

\begin{figure}[t]
  \centering
  \includegraphics[width=0.4\textwidth]{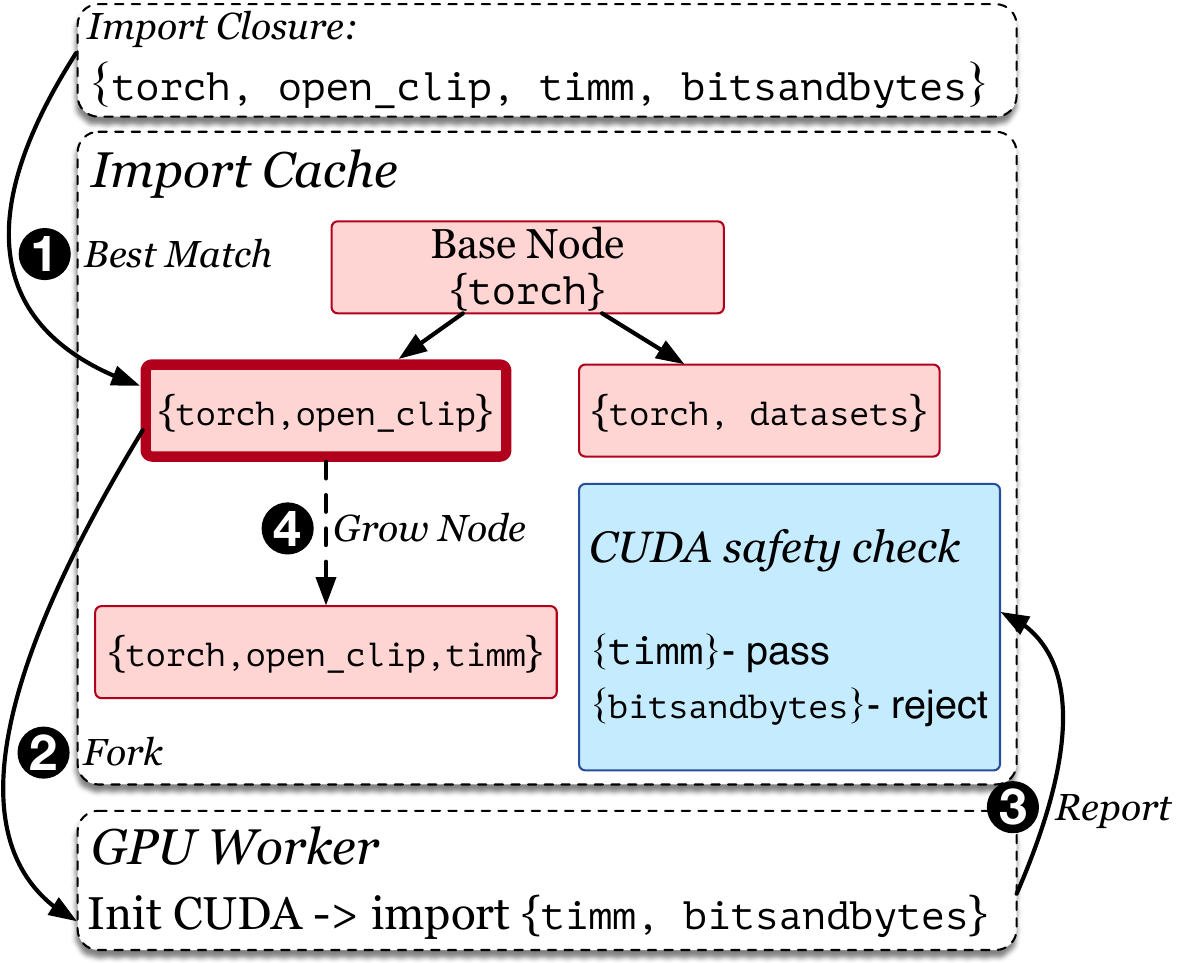}
  \caption{The GPU-aware import cache. Cache nodes are live CUDA-clean
  zygote processes; the base node holds the common head and child nodes
  extend it with tail packages. For a cell whose import closure is
  \{\texttt{torch}, \texttt{open\_clip}, \texttt{timm},
  \texttt{bitsandbytes}\}, the worker forks from the best-matching node
  (\protect\circled{1}, \protect\circled{2}) and imports the rest after
  the fork. \texttt{timm} passes the CUDA-safety check, so the cache
  grows a node for it asynchronously (\protect\circled{3},
  \protect\circled{4}); \texttt{bitsandbytes} touches CUDA and stays
  out of the cache.}
  \label{fig:importcache}
\end{figure}

\phead{Cache hierarchy.}
\sys{} implements the import cache as a hierarchy of cache nodes
(\Cref{fig:importcache}).
Each cache node is a live CUDA-clean zygote process
with a set of loaded modules. The base node holds the common packages from the head of the distribution, and child nodes extend it with tail packages.
For a GPU cell, \sys{} chooses the cache
node whose loaded modules best match the import closure of the cell (\circled{1}). The worker
forks from that node and inherits its loaded modules (\circled{2}).
After the fork, the worker
initializes CUDA and then imports any remaining packages.

\begin{figure*}[t]
  \centering
  \includegraphics[width=0.95\textwidth]{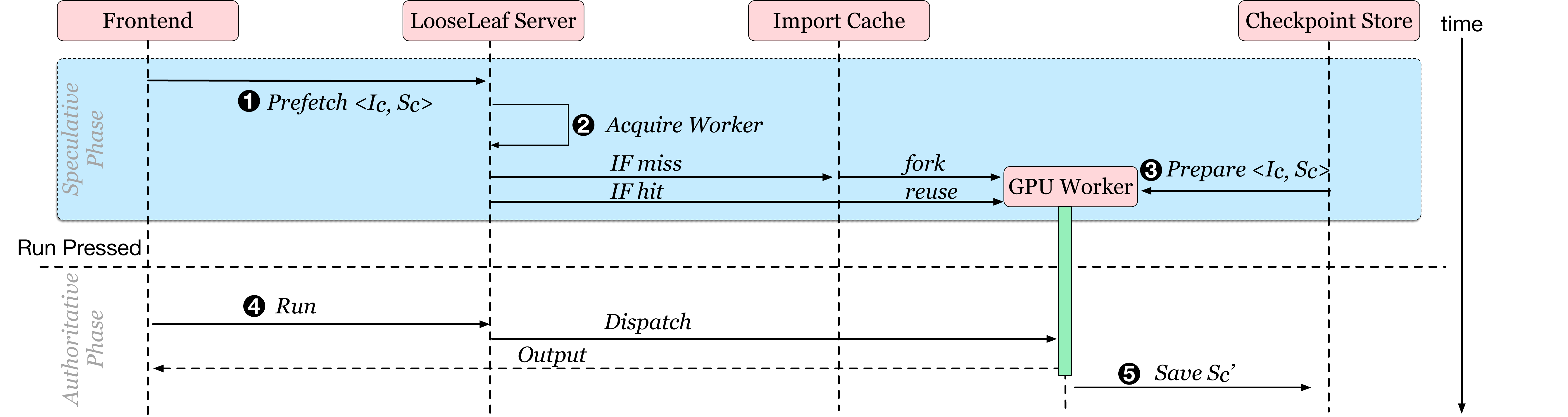}
  \caption{ During editing, the frontend sends a prefetch request that carries the
  current closures (\protect\circled{1}). The server acquires a worker for the session
  (\protect\circled{2}). On a miss, the server forks a new worker from the import cache. On a hit, it reuses the warm worker of the session. The worker imports the remaining packages and restores the state closure from the
  checkpoint store (\protect\circled{3}). When the user presses Run (\protect\circled{4}), the
  worker completes any remaining preparation and executes the cell. After
  execution, the worker saves the updated state to the checkpoint store
  (\protect\circled{5}). The worker is later reclaimed and the state persists in the
  store.}
  \label{fig:lifecycle}
\end{figure*}

\phead{Growing the cache.}
When a cell introduces imports that are absent from the selected cache node, the
worker reports these imports to the import cache before executing the cell (\circled{3}).
The import cache then grows a new node asynchronously (\circled{4}) by forking the
selected node and importing the packages into it, so that later workers
inherit them. Because this growth is asynchronous, it does not delay
the execution of the cell.
To keep every node CUDA-clean, \sys{} applies a CUDA-safety check and
admits only packages whose import does not touch CUDA. A package that
touches CUDA stays out of the cache, and each worker imports it after
the fork, which is safe because the worker has already initialized
CUDA. Each host caches a bounded number of nodes and evicts the least
recently used leaf node, and the base node stays resident to keep the
common head of the distribution loaded.

\subsection{Bounded Edit-Time Speculation}
\label{sec:design-spec}

The two preceding mechanisms lower the cost of creating a GPU worker.
The remaining question is when and where to create it. If the system
waits until the user presses Run, the preparation latency falls on the
critical path. Notebook editing, however, provides a short preparation
window before execution (\cref{sec:obs3}). \sys{} exploits this window
by preparing GPU workers while the user edits a cell, and the DAG
daemon keeps the def-use graph current with each edit for the
preparation steps below.

\phead{Routing cells to the GPU path.}
The frontend first needs to decide whether the candidate cell should use the GPU
path. Running every cell on the GPU path would waste GPU resources, since many
notebook cells are CPU-only. This decision can rely on static package
dependencies, because notebook cells are Python code,
and GPU work is invoked through Python packages that bind to GPU
libraries such as PyTorch~\cite{pytorch_nips19} and related AI/ML frameworks.
\sys{} therefore identifies GPU cells by their \emph{package
use}. From the def-use graph, \sys{} computes the \emph{effective
import set} of each cell. The set contains every imported package
that is reachable from the references in the top-level statements of
the cell. In \Cref{fig:dag-closure}, the set of Cell~5 contains
\importhl{\texttt{transformers}} through the reference to
\texttt{Trainer} and \importhl{\texttt{torch}} through the call in
\statehl{\texttt{compute\_map3}}, whereas the set of Cell~4 is empty
because its top-level statement only defines a function. \sys{}
routes a cell to the GPU path only when its effective import set
contains a package from a list of GPU framework packages.
\sys{} builds this list both automatically and manually. A package
installed on the worker hosts enters the list when its dependencies
include the CUDA runtime, and users can add further packages.

\begin{algorithm}[t]
\caption{Per-cell worker acquisition. $L$ is the GreedyDual priority defined under \emph{Bounding GPU residency}.}
\label{alg:worker}
\KwIn{session $s$, cell $c$, per-host warm worker cache $W$}
\uIf{$s$ has a warm worker $w$ in $W$}{
  restore missing state into $w$ \tcp*{hit}
}
\Else(\tcp*[f]{miss}){
  $h \gets$ most-loaded host with sufficient memory and a free GPU slot\;
  \If{no such host exists}{
    $h \gets$ most-loaded host with an evictable idle worker\;
    evict the worker of session $\arg\min_{s'} L(s')$ from $W[h]$ \tcp*{GreedyDual}
  }
  $w \gets$ fork from import cache on $h$\;
  \If{$h \neq$ previous host of $s$}{
    migrate state of $s$ to $h$\;
  }
  restore state closure $S_c$ into $w$\;
}
\end{algorithm}

\phead{Prefetch and worker acquisition.}
For a GPU-classified candidate cell, the frontend sends the current closures to the server as a prefetch request.
The server then acquires a GPU worker for the session. If the session
already has a warm worker, the server reuses it and restores only the
missing or outdated variables, which constitutes a \emph{hit}.
Otherwise, the server creates a worker on a host chosen by
consolidation (described below), which constitutes a \emph{miss}. The
new worker forks from the import cache (\cref{sec:design-env}) and
restores the state closure from the checkpoint store
(\cref{sec:design-state}). When the chosen host differs from the
previous host of the session, the session state migrates first.
Algorithm~\ref{alg:worker} summarizes the procedure.

\phead{Bounding GPU residency.}
Edit-time preparation also creates a resource-control problem. A prefetched
worker occupies a GPU device, so the system must decide how long the worker
should remain resident. A fixed timeout provides no stable choice:
a short timeout can discard useful preparation, while a long timeout lets a client hold a GPU with cheap prefetch requests.

Instead of a fixed timeout, \sys{} manages idle warm workers as a
cache. An idle worker can be seen as a cache entry: a later cell of
the same session can hit the worker and skip its preparation. \sys{}
manages the cache with GreedyDual~\cite{greedydual_algorithmica94},
so eviction weighs recency together with restore cost. For a session
$s$, let $R_s$ be the GPU state held by its warm worker. \sys{}
computes
\[
  c(s) = \sum_{r \in R_s} \mathrm{bytes}(r), \qquad
  L(s) = H_s + c(s),
\]
where $H_s$ is the value of the host's clock when $R_s$ last changed;
the clock never decreases. Under GPU pressure, \sys{} first evicts
idle workers whose cell has completed. When no such worker exists, it
evicts a speculatively prepared worker. Within each class, it evicts
the worker with the smallest $L(s)$ and advances the clock to that
value.

The policy bounds the residency that any client can obtain. Because
$H_s$ never exceeds the clock, $L(s)$ lies at most $c(s)$ above it,
so the worker is evicted by the time the clock climbs that far unless
$L(s)$ rises, and $L(s)$ changes only with $R_s$. Apart from executing
a cell, a client changes $R_s$ only through a prefetch, which loads
variables that the worker does not yet hold, and every such variable
comes from an earlier execution of the session. A repeated prefetch
for the same closure therefore leaves $L(s)$ unchanged while the clock
climbs past it. The residency of a session is thus bounded by the
state bytes that its cells have already computed: cheap prefetch
requests cannot hold a GPU, and an adversary obtains no more
residency than an honest session that computed the same state.

\phead{Worker placement and cluster provisioning.}
\sys{} places new workers by consolidation. Among the hosts with
enough memory for the session state, the server picks the most loaded
host with a free GPU slot, and when no such host exists it evicts an
idle worker on the most loaded host under the policy above. A
background controller scales the number of powered hosts with a
target-utilization policy~\cite{k8shpa_web} and square-root staffing
headroom~\cite{halfinwhitt_opres81}, and it powers a host off only
after its warm workers are evicted and the resident state migrates to
another host.

\phead{Life of a GPU cell.}
\Cref{fig:lifecycle} traces the protocol between the frontend and
the server. The protocol has a speculative phase during editing and
an authoritative phase after the user presses Run. During editing,
the prefetch request carries the current closures of the cell, so the
preparation of the worker overlaps with the edit window
(\cref{sec:obs3}). The speculation is safe because a prefetch only
loads packages and variables and never executes the cell. The
prepared worker may still become stale before the run. The run
request therefore carries the closures computed from the submitted
code, and the worker checks the versions of the restored variables
(\cref{sec:design-state}), refreshes the outdated ones from the
checkpoint store, and completes any remaining preparation before
execution. A stale prefetch thus costs only the uncovered preparation
time and never affects correctness. After execution, the worker saves
the updated state and stays warm under the eviction policy above.

\section{Evaluation}
\label{sec:evaluation}

\subsection{Experimental Setup}
\label{sec:eval-setup}

\begin{figure*}[t]
  \centering
  \includegraphics[width=0.95\textwidth]{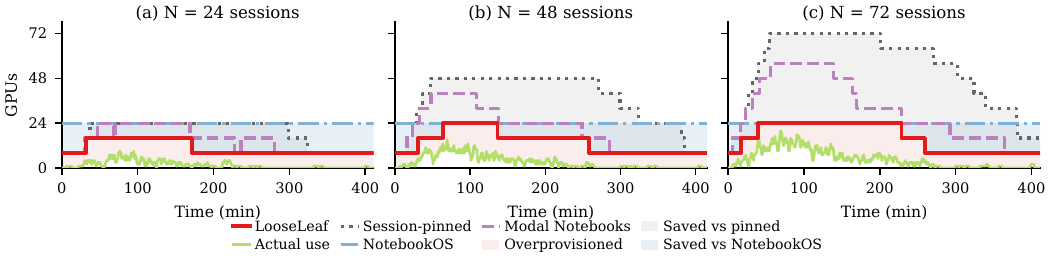}
  \caption{Provisioned GPUs over time for 24, 48, and 72 concurrent
  sessions, against the GPUs actually executing cells. Provisioning
  acts at the whole-host granularity of eight GPUs. The shaded regions
  mark the GPU time that \sys{} saves against each baseline and the
  headroom that \sys{} holds above actual use.}
  \label{fig:eval-gpu}
\end{figure*}

\phead{Implementation.}
We implemented \sys{} for both Jupyter and
marimo~\cite{marimo_web}. In both, the frontend maintains the def-use
graph during editing and runs the closure computation and GPU
labeling inside a Pyodide~\cite{pyodide_web} web worker, and the two
frontends share the same server and workers.
The import cache is a
per-host process tree with LRU eviction capped at 50 nodes. The
checkpoint store keeps metadata in Redis~\cite{redis_web} and
payload bytes in a hugetlbfs pool managed by
dlmalloc~\cite{dlmalloc_web}. Payloads move across hosts through
the zero-copy interface of Linux~\cite{msgzerocopy_netdev17}, and
structure blobs synchronize through the object store of
Ray~\cite{ray_osdi18}. The server exports the demand signal to a
standard cluster autoscaler~\cite{k8shpa_web}.

\phead{Testbed.}
All measurements run on
a cluster of AWS \texttt{p4d.24xlarge} instances with 8
NVIDIA A100-40\,GB GPUs, 96 vCPUs, and 1,152\,GiB of RAM. Each node
reserves a 512\,GiB hugepage arena that backs the checkpoint payload
pool (\cref{sec:design-state}).

\begin{table}[t]
  \centering
  \small
  \caption{Evaluation workload: 24 Kaggle notebooks.
  Each entry shows the realized notebook count,
  which equals the corpus-proportional target.}
  \label{tab:workload}
  \setlength{\tabcolsep}{4pt}
  \scalebox{0.92}{
  \begin{tabular}{l r r r r}
    \toprule
    \textbf{Task type} & \textbf{small} & \textbf{medium} & \textbf{large} & \textbf{Total} \\
                       & ($\le$300M)    & (300M--1B)      & ($\ge$1B)       &                \\
    \midrule
    training-only      & 4 & 0 & 0 & 4 \\
    training+inference & 9 & 1 & 1 & 11 \\
    PEFT+inference     & 0 & 0 & 3 & 3 \\
    inference-only     & 2 & 0 & 4 & 6 \\
    \midrule
    \textbf{Total}     & \textbf{15} & \textbf{1} & \textbf{8} & \textbf{24} \\
    \bottomrule
  \end{tabular}
  } %
\end{table}

\phead{Workload.}
We evaluate on 24 Kaggle notebooks sampled from the 2024--2025
competition corpus. The sample follows the task-type and model-size
distribution of the corpus by Hamilton apportionment~\cite{fairrep_brookings01}
(Table~\ref{tab:workload}). The notebooks cover training, training with
inference, parameter-efficient fine-tuning, and inference-only
pipelines. The models range from CNNs below 300\,M parameters to
14\,B-parameter LLMs.
Inter-cell arrival time is drawn from the Adobe notebook traces~\cite{notebookos_asplos26}. The edit window of each cell is drawn from the paste-to-run distribution measured in DSPM~\cite{dspm_chi26} (\cref{sec:obs3}).

\phead{Baselines.}
We compare \sys{} against four baselines:
\begin{itemize}[leftmargin=*,noitemsep]

\item \emph{Session-pinned
reservation} is the model of Jupyter, Colab, and Kaggle: a session
holds a GPU from its arrival until the session ends. It defines the
latency ideal, since every cell finds a
warm kernel. It also defines the provisioning ceiling, since the
reserved GPU spans every idle gap between cells. We model it
charitably: sessions are packed best-fit onto 8-GPU hosts, and a host
powers off as soon as it becomes empty.

\item \emph{NotebookOS}~\cite{notebookos_asplos26} keeps three persistent kernel
replicas for each session, so an arriving GPU cell always finds a
ready kernel to serve it. When a cell runs, NotebookOS assigns a free
GPU to the replica that executes it. The replicas hold the session
state in CPU memory, so a GPU cell must first move the accumulated
model state onto the GPU before it executes. After a cell updates a
model, the executing replica serializes the updated state and
synchronizes the other replicas. This synchronization runs in the idle
gap between cells. When it takes longer than the gap, the next cell
waits for the remainder.

\item \emph{Modal}~\cite{modal_web} is a production serverless GPU platform
that also provides a serverless notebook service.
It keeps one kernel per
session and reclaims the kernel after a default 10-minute idle timeout. The reclamation discards the session state.
A later cell must reboot the notebook server and rerun all upstream cells.

\item \emph{Naive serverless} runs each GPU cell as a plain
serverless GPU function: it starts a fresh container, initializes CUDA,
cold-imports its frameworks, and transfers the whole live namespace
in; after execution, it writes the whole namespace back. This
configuration equals \sys{} with the dependency graph, the checkpoint
store, and the import cache all disabled.

\end{itemize}

\phead{Metrics.}
For provisioning, we report \emph{GPU-hours}, the time integral of
provisioned GPUs at the whole-host granularity of eight GPUs, and the
peak GPU count; both are compared with the GPUs actually executing
cells. For interactivity, we report \emph{user-visible latency}, from
pressing Run until the cell begins execution, excluding cell execution
time. Following \cref{sec:intro}, we use 1\,s as the
\emph{interactive-start threshold} and assess p95 interactivity by whether
at least 95\% of cell runs begin within this threshold. To explain these
results, we report \emph{preparation latency} (\cref{sec:closures}), the
time required to make a GPU worker ready, relative to the edit window, and
\emph{save time}, the post-execution checkpoint cost, relative to the
inter-cell interval. Save time also determines how quickly the worker can
be reclaimed.
\subsection{GPU Provisioning at Scale}
\label{sec:eval-gpu}

Figure~\ref{fig:eval-gpu} reports the GPUs each baseline holds as the
load grows from 24 to 72 concurrent sessions. Provisioning acts at
whole-host granularity, so powering one host on adds eight GPUs. The
green line gives the ground truth, the number of GPUs executing cells
at each instant. Its peak stays at 9, 15, and 20 GPUs across the three
loads, because sessions spend most of their time thinking and editing
between cells.

\sys{} binds a GPU to a cell rather than to a session. Between cells,
it holds only warm workers, and a warm worker is evictable at any
time because the session state persists in the checkpoint store.
The worker-management policy (\cref{sec:design-spec}) concentrates these workers on a few hosts.
Consolidation packs new workers onto the most-loaded hosts, so
lightly loaded hosts drain. Square-root staffing boots hosts ahead of
the arrival burst, and a drained host that stays idle is powered off.
These policies appear as the steps up to 16 or 24 GPUs
and back down to 8. As the load triples, the peak provisioning of
\sys{} grows only from 16 to 24 GPUs.

Session-pinned reservation scales with the number of sessions rather
than with this demand. Every session holds one GPU from arrival until
the session ends, so the reserved peak follows the session count, at
24, 48, and 72 GPUs. \sys{} provisions 43\%, 63\%, and 70\% fewer
GPU-hours across the three loads. The saving grows with the load,
because the reservation cost grows linearly while \sys{} tracks actual
use.

NotebookOS holds the same 24 GPUs at every load. The three replicas
of a session must reside on three distinct hosts, so three powered
hosts form its floor. The replicas are resident in CPU memory and
block scale-in, so the host count never falls. At 24 sessions
most of this floor sits idle, and \sys{} holds 55\% fewer GPU-hours.
At 72 sessions the demand grows into the floor, and the saving narrows
to 32\%.

Modal bounds each session reservation with the 10-minute idle
timeout. A GPU returns only when a pause exceeds the timeout, so Modal
recovers the long pauses and holds the GPU through the rest of the
think time. Its peak reaches 24, 40, and 56 GPUs, and \sys{} provisions
29\%, 33\%, and 47\% fewer GPU-hours. A shorter timeout would recover
more idle time, but every expiry discards the session state, and the
next cell pays the recomputation that \cref{sec:eval-latency} reports.
\sys{} avoids this tradeoff. The session state persists in the
checkpoint store, so a warm worker is evictable.

Among the four baselines, only the provisioning of \sys{} follows the
actual use. Session-pinned reservation and Modal provision by the
session count, and NotebookOS provisions a static floor. The remaining
headroom of \sys{} above the ground truth comes from the whole-host
granularity, the square-root staffing, and the warm workers held
between cells. Next, \cref{sec:eval-latency} shows that this saving does not come at the cost of interactivity.
\subsection{End-to-End Interactive Latency}
\label{sec:eval-latency}

\begin{figure}[t]
  \centering
  \includegraphics[width=0.95\columnwidth]{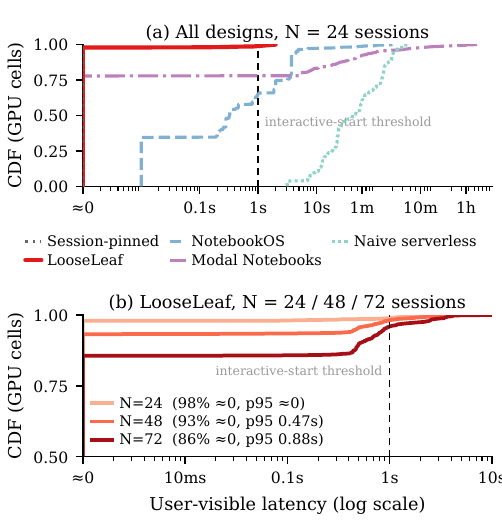}
  \caption{User-visible latency over the GPU cells (log scale).
  (a)~All designs with 24 concurrent sessions.
  The curve of \sys{} coincides with the session-pinned
  ideal at the left edge. (b)~\sys{} as the load grows to 48 and 72
  sessions. The vertical dashed line marks the
  one-second interactive-start threshold.}
  \label{fig:eval-latency}
\end{figure}

Figure~\ref{fig:eval-latency}(a) reports the distribution of
user-visible latency over the GPU cells, with 24 concurrent sessions
under the AI-assisted edit window. In every design, the latency of a
GPU cell comes from two sources. The first source is preparation that
begins only after the user presses Run. The second source is
unfinished work of the previous cell, which the new run must wait
for. The five designs differ in how much of each source they expose
to the user.

\sys{} removes the first source by preparing during editing. The
figure shows that the uncovered part is negligible: 98\% of GPU cells
begin executing immediately after Run. The exposed latency stays below
$1.1$\,s at p99 and $2.0$\,s at the worst case. 98.9\% of
GPU-cell runs start within the interactive-start threshold. Two mechanisms keep the preparation within the window.
For 79\% of GPU cells, the session finds its warm worker in the worker
cache, and the worker restores only the variables that changed since
the last run. The remaining cells fork a fresh worker from the import
cache and restore the state closure from the checkpoint store. This
cold preparation is itself small, with a median below one second
(\cref{sec:eval-prep}).
The residual tail comes from the cells whose edit window is
shorter than this cold preparation. The user runs such a cell before
the fork and restore complete, and the unfinished remainder is
exposed. The second source is also absent in \sys{}: a
GPU worker saves the updated variables immediately after execution,
and the incremental save completes long before the next run
(\cref{sec:eval-save}).

NotebookOS suffers from both sources. Its preparation is reactive: no
transfer can start before Run, and without a closure the executor
moves every model the session has accumulated, so the exposed
transfer grows with the session. The median is $0.33$\,s and p95 is
$4.4$\,s. The second source dominates the tail. After each cell that
updates a model, NotebookOS serializes the full model state and
writes it out between cells.
The save of \sys{}, in contrast, is incremental. Dirty
tracking refreshes only the changed regions of the updated model, and the copy runs as pinned DMA into the checkpoint store (\cref{sec:design-state}).
On the notebooks that accumulate tens of
gigabytes of state, this synchronization outlasts the gap
between two runs, and the next run waits for the remainder. The tail
reaches $63$\,s at p99 and $194$\,s at the maximum. Only
65\% of its runs start within the threshold.

Modal starts a GPU cell immediately while its container is warm,
which covers 78\% of GPU cells. After an idle shutdown, the next run
reboots the endpoint and reruns all upstream cells. This
recomputation takes $46$\,s at p90 and exceeds an hour in the tail.
Naive serverless pays both sources on every cell.
After Run, it starts a fresh container, initializes CUDA, imports the
frameworks from scratch, and transfers the whole live namespace into
the worker, for a median of $41$\,s. After execution, it writes the
whole namespace back, and this write-back delays the next run in the
same way. No run of naive serverless starts within the
threshold. Each of these costs corresponds to one mechanism of \sys{}:
the import cache removes the container start and the imports, the
state closure shrinks the transfer, and the incremental save shrinks
the write-back.
\sys{} is the only design that keeps both sources near
zero while keeping the GPU reclaimable between cells.

The latency stays low as the load grows.
As the sessions
triple, the peak GPU count that \sys{} provisions grows only from 16
to 24 (\cref{sec:eval-gpu}). Three times the sessions share this
nearly unchanged capacity, so warm workers are evicted more often and the
hit rate falls from 79\% to 48\%. More GPU cells therefore prepare on
a fresh worker. Figure~\ref{fig:eval-latency}(b) shows that the cost
stays small. At 72 sessions, 86\% of GPU cells still start
immediately, and p95 stays at $0.88$\,s. Even at 72 sessions,
95.9\% of runs start within the interactive-start threshold. The residual tail is bounded
by the time to power on a node, which a demand spike occasionally pays
before the autoscaler catches up. The GPU savings of
\cref{sec:eval-gpu} therefore do not come at the cost of
interactivity.

\subsection{Why Preparation Fits the Edit Window}
\label{sec:eval-prep}

\begin{figure}[t]
  \centering
  \includegraphics[width=0.95\columnwidth]{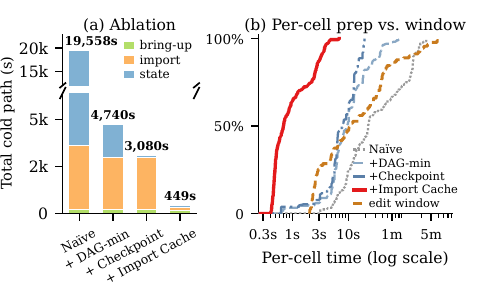}
  \caption{(a)~Cold-path preparation summed over the
  workload as the mechanisms are enabled cumulatively, split into
  worker bring-up, module import, and state transfer/restore.
  (b)~The per-cell distribution of the cold preparation at each
  step, against the distribution of the edit window.}
  \label{fig:eval-ablation}
\end{figure}

\begin{table}[t]
  \centering
  \small
  \caption{ Checkpoint and restore of four model states from
  the workload. The
    full save is the first save of the state and writes it whole. The
    incremental save is issued by the cells after the first and
    refreshes only the changed regions.}
  \label{tab:microbench}
  \setlength{\tabcolsep}{3.5pt}
  \scalebox{0.92}{
  \begin{tabular}{l r r r r}
    \toprule
    \textbf{Model} & \textbf{State size} & \textbf{Full save} &
    \textbf{Restore} & \textbf{Incr.\ save} \\
    \midrule
    DINOv3-L   & 1.2\,GB  & 0.18\,s & 0.29\,s & 29\,ms \\
    ESMC-300M  & 2.7\,GB  & 0.41\,s & 0.48\,s & 47\,ms \\
    Mistral-7B & 14.5\,GB & 1.57\,s & 2.03\,s & 66\,ms \\
    Qwen3-8B   & 15.1\,GB & 1.63\,s & 2.04\,s & 19\,ms \\
    \bottomrule
  \end{tabular}
  } %

\end{table}

\Cref{sec:eval-latency} shows that both latency sources stay near
zero for \sys{}. The rest of the evaluation explains why. This
section covers the first source, preparation. \sys{} performs it
during editing, so it stays invisible only when it finishes within
the edit window. The measurements release the GPU worker
after every cell, so each GPU cell prepares on a fresh worker and
pays the whole cold path.

Figure~\ref{fig:eval-ablation}(a) enables the three mechanisms one after
another and reports the cold preparation of the whole workload at
each step. The naive configuration spends $19{,}558$\,s, and
$15{,}951$\,s of it moves the whole live namespace. Dependency-guided
minimization restores only the state closure of each cell and imports
only the packages the cell needs. It cuts the state term by
$8.9\times$ and the total by $4.1\times$. Enabling the checkpoint store moves the
large state payloads at high speed, and the state term falls to
$124$\,s. The import cache removes the
repeated imports, so the import term falls $18\times$ and the total reaches $449$\,s. The three steps shrink the cold path by $44\times$.

Table~\ref{tab:microbench} reports these costs on single model
states. A restore is dominated by the DMA: reloading a 14--15\,GB
model takes about two seconds at 7.6--8.1\,GB/s. The
transfer runs over the host-to-GPU PCIe path of the instance, whose
measured pinned-memory bandwidth peak is 12.4\,GB/s.
The remaining time reconstructs the Python object structure.

With all three mechanisms enabled, the cold preparation of a
cell has a median of $0.71$\,s and a p95 of $3.8$\,s.
Figure~\ref{fig:eval-ablation}(b) places the per-cell
preparation distribution of each step against the distribution of
the edit window, which every cell faces alike. Under the naive
configuration most preparations outlast most windows, so the
preparation cannot hide in the window. Each mechanism shifts the
preparation distribution left.
In the full configuration, preparation has a maximum of 6.9\,s, while the edit-window
distribution has a median of 10.7\,s, indicating substantial opportunity to
hide preparation during editing. Note it does not imply that every cold
preparation is fully hidden: the figure does not pair each cell's preparation
time with its realized edit window, and a long preparation may coincide with
a shorter window.

\subsection{Comparison with Transparent GPU Checkpoint/Restore}
\label{sec:eval-cr}

\begin{figure}[t]
  \begin{minipage}{\columnwidth}
    \centering
    \includegraphics[width=0.85\columnwidth]{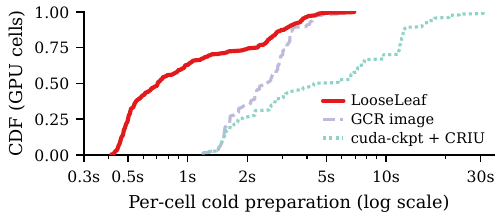}
    \caption{Comparing \sys{}'s restore cost with state-of-the-art transparent checkpoint/restore approaches.}
    \label{fig:restore-cdf}
  \end{minipage}
  \par\vspace{8pt}
  \begin{minipage}{\columnwidth}
    \centering
    \includegraphics[width=0.85\columnwidth]{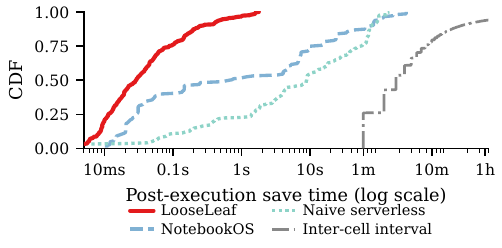}
    \caption{Post-execution save time of each design against the
    inter-cell interval of the production trace.}
    \label{fig:eval-save}
  \end{minipage}
\end{figure}

Dependency-guided restore is not the only way to prepare a GPU
worker. Transparent checkpoint/restore offers an alternative.
CRIU~\cite{criu_web} with cuda-checkpoint~\cite{cudackpt_web} saves
the kernel process together with its GPU memory and restores both on
demand. GCR~\cite{gcr_fast26} is a recent GPU checkpoint system that restores
GPU memory at close to the PCIe bandwidth. In both approaches,
restoring a GPU cell first starts the container. The restore then
loads the process memory and the GPU memory that the session has
accumulated. We estimate this per-cell restore time from the measured
memory sizes and transfer rates.
Figure~\ref{fig:restore-cdf} shows the per-cell preparation time. The
median preparation of \sys{} is 0.71\,s. The GCR-based restore has a
median of 2.50\,s. The CRIU-based restore has a median of 4.38\,s. The
difference comes from the amount of state that each side restores. The
checkpoint-based approaches restore the entire session. They also pay
a fixed cost for the container start before any data moves. \sys{}
restores only the state closure and the import closure of the target
cell. The checkpoint-based approaches are faster
only at the tail. The slowest cell of \sys{} takes 6.9\,s. Most of
this time imports \texttt{bitsandbytes}, which is not a CUDA-safe
package and cannot be optimized by the import cache.

\subsection{Why the Save Fits the Interval}
\label{sec:eval-save}

The second latency source of \cref{sec:eval-latency} is the save of
the previous cell. After a cell executes, \sys{} saves the updated
state to the checkpoint store, and this save must finish before the
next run needs the state. The save of \sys{} is incremental at two
levels (\cref{sec:design-state}): it saves only the variables that
the cell updated, and within a GPU tensor, dirty tracking refreshes
only the 2\,MB regions that changed since the previous checkpoint.
An incremental save scans the state for the changed regions
and transfers only those regions.
Table~\ref{tab:microbench} shows
the effect: the full save of a 15\,GB state takes 1.6\,s at
9.6\,GB/s of D2H DMA,
and the
incremental save of a later cell takes tens of milliseconds.
NotebookOS saves the complete
model and optimizer state on every synchronization. Naive serverless
saves the whole live namespace after every cell.

Figure~\ref{fig:eval-save} compares the three saves against the
interval they must fit into, the time between consecutive runs in the
production trace. Session-pinned reservation and Modal keep state
only in the running kernel and write nothing back. The incremental save of \sys{} has a median of
$29$\,ms and never exceeds $1.8$\,s. The smallest interval in the
trace is $60$\,s, so no save of \sys{} delays the next run in this
workload. The saves of NotebookOS and naive serverless cross into the
interval distribution. The save of NotebookOS reaches $109$\,s at p95
and $254$\,s at the maximum, which is $177\times$ the p95 of \sys{},
and it overflows 4.6\% of the trace intervals. The save of naive
serverless overflows 6.8\%. These overflowing saves are the source of
the latency tails in \cref{sec:eval-latency}.

A fast save also matters for provisioning. A warm worker is evictable
only after its save completes, because eviction must not lose state.
The save of \sys{} finishes within seconds of every cell, so every
warm worker becomes evictable almost immediately. This enables the
demand-following provisioning of \cref{sec:eval-gpu}.

\section{Related Work}
\label{sec:related}

NotebookOS~\cite{notebookos_asplos26} allocates GPUs
to cells and is the
strongest baseline in \cref{sec:evaluation}. This section covers the
remaining related work.

\phead{Notebook state management.}
ElasticNotebook~\cite{elasticnotebook_vldb23} checkpoints session
state for live migration and chooses between storing and recomputing
each variable. Kishu~\cite{kishu_vldb24} versions the session state
and restores a past state through incremental checkout.
Chipmink~\cite{chipmink_vldb25} partitions the object graph into
cost-based pods and persists only the pods that changed.
nbsafety~\cite{nbsafety_vldb21} traces lineage across cells to warn
about stale results. These systems manage the CPU state of a session
that runs in one long-lived kernel, and their checkpoints are read
back at a failure, a checkout, or a migration. \sys{} extends this
line of work to GPU-resident state that moves between workers. The
store sits on the execution path. Every cell restores its closure from
the store and saves its updated variables back
(\cref{sec:design-state}).

\phead{Transparent GPU checkpoint and restore.}
CRIU~\cite{criu_web} with cuda-checkpoint~\cite{cudackpt_web} snapshots
a process together with its GPU state. PhoenixOS~\cite{phoenixos_sosp25}
checkpoints GPU applications concurrently with their execution.
GCR~\cite{gcr_fast26} restores GPU memory near the PCIe bandwidth
through incremental checkpoints of dirty regions. These systems are
transparent. They capture the whole address space and cannot tell which
part a cell needs, so a restore always carries the full session. \sys{}
restores only the state closure of the target cell and preserves object
sharing across the restored variables. The difference in preparation
time is quantified in \cref{sec:eval-cr}.

\phead{Serverless data systems and cold start.}
Interactive data systems increasingly run on serverless
platforms. Starling~\cite{starling_sigmod20} and
Lambada~\cite{lambada_sigmod20} run analytics on cloud functions,
Cloudburst~\cite{cloudburst_vldb20} co-locates a cache with function
executors, Vexless~\cite{vexless_sigmod24} keeps vector indexes in
stateful functions, and Azure SQL Database Serverless resumes paused
databases ahead of the next request~\cite{azureserverless_sigmod24}.
General platforms reduce the start delay of any function with prebuilt
snapshots~\cite{catalyzer_asplos20, reap_asplos21}, predicted
keep-alive windows~\cite{serverlesswild_atc20}, and snapshots of an
initialized CUDA context~\cite{modalsnapshot_web}. These systems
cannot determine the state that the next request requires until the
request arrives, so they either hold the complete state ready or
obtain it on the critical path. In a notebook, the code of the next
cell is available before execution, and \sys{} prepares only the
closure of that cell during editing, before the user requests
execution.

\section{Conclusion}

This paper presented \sys{}, a novel computational notebook platform with serverless GPUs.
A notebook cell needs only a
small part of the session state, which can be derived from the code
while the user edits the cell. \sys{} captures this part as the
cell's state and import closures. It prepares a GPU worker with these
closures during editing and saves the updated state within seconds
after execution. GPUs are therefore allocated to cells on demand
rather than pinned to sessions, reducing idle GPU time
and improving GPU utilization.

\bibliographystyle{ACM-Reference-Format}
\bibliography{references}

\end{document}